# Comments on structural types of toroidal carbon nanotubes

Chern Chuang,[a] Yuan-Chia Fan,[a] and Bih-Yaw Jin*[a]

In a recent article[1], Beuerle *et al.* identified eight structural types of high-symmetry achiral toroidal carbon nanotubes (TCNTs) based on two of our previous papers on the same subject[2-3]. One can rationalize these eight types of TCNTs in the following way: A TCNT can be viewed as a polygonal torus with coaxial outer- and inner-rims that can be either prism or antiprism, but not independently. If inner-rim is a prism (antiprism), then outer-rim must be a prism (antiprism), and vice versa. Moreover, there are two possible ways to arrange inner-rim with respect to outer-rim, parallel and anti-parallel, by which we mean the inner- and outer-rim are in the eclipsed and staggered arrangement, respectively. Thus, there are four possible shapes of polygonal TCNTs (Table 1). For each of these four shapes, there are two possible ways to tile graphene sheet onto its polygonal surface achirally, therefore, there are in total eight structural types of TCNTs. Beuerle et al. further described how these eight structural types of TCNTs can be transformed in a cyclic way by employing the two kinds of transformations, the horizontal shifting (HS) and the concerted Stone-Wales transformations (cSWT).

Table 1. The four basic shapes of TCNTs.

| I | II | III | IV |
|---|---|---|---|
| 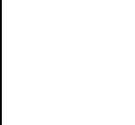 | 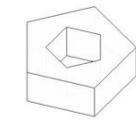 | 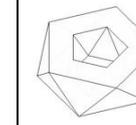 | 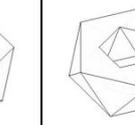 |

Here, in this correspondence, we would like to point out that, while the shapes of these eight structural types are correctly stated, the transformations that bring them into each other are more subtle than that described inprecisely by the authors of [1]. Particularly, we believe that they have mistaken the cSWT, which we called the generalized Stone-Wales transformation (gSWT), as another type of transformation we introduced, *i.e.* the rim rotation (RR). This is because the gSWT can only be applied to the TCNTs belonging to the $D_{nd}$ point group and the resulting TCNTs still have well-defined latitude coordinates. And only in this particular case, the gSWT and RR are equivalent to each other. One cannot apply the gSWT to $D_{nh}$-TCNTs in the same way as we applied it to the $D_{nh}$-TCNTs. Moreover, in our second paper, we pointed out that, for certain $D_{nh}$-TCNTs, one can still apply the gSWT in a different way to get high-symmetry TCNTs that do not have latitude coordinates defined[3]. A complete list of these five extra structural types will be given in the second part of this correspondence. Thus, we believe that it is important to make a clear distinction between these two types of transformations because they have different effects when operating on the parent TCNTs. To begin with, we will make a brief review on these three types of transformations, i.e. RR, gSWT, and HS.

[a] Mr. C. Chuang, Mr.Y.-C. Fan, Prof. B.-Y. Jin
Department of Chemistry, Center for Theoretical Sciences, and Center for Quantum Science and Engineering
National Taiwan University
No. 1, Roosevelt Road Sec. 4, Taipei, Taiwan
Fax: (+)
E-mail: byjin@ntu.edu.tw

**Rim rotation for $D_{nh}$-TCNTs:** In view of these polygonal tori whose vertices are the nonhexagonal rings, RRs are the transformations that alter the relative positions of the vertices. Applying a RR on a (parallel)-TCNT, prismatic ($D_{nh}$) or antiprismatic ($D_{nd}$), will cause the inner part of the polygonal tori to become staggered (antiparallel) with respect to the outer part. We can further distinguish RR to be inner RR and outer RR, denoted as IRR and ORR hereafter for convenience, depending on whether it changes the local atomic constituents and connectivities of the inner and the outer part of the parent TCNT (Figure 1).

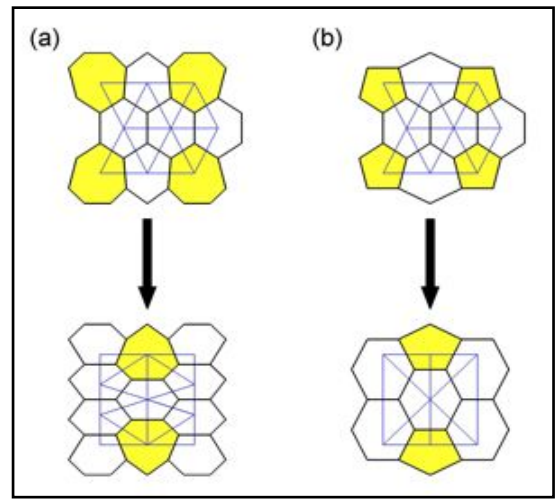

Figure 1. Illustration of rim rotation in $D_{nh}$ TCNTs. The connectivity of the neighboring atoms are shown in black edges and nonhexagons are shaded yellow. The dual representation of the affected atoms (blue triangles) are marked with blue lines.

Here we have cut the rectangular parts of a rotational unit cell of prismatic polygonal TCNT with the chiral vector (1,0). The patch with heptagons (located in the inner-rim of the TCNT) is shown in Figure 1(a) and one with pentagons (located in the outer-rim of the TCNT) in Figure 1(b). We shall stick to this convention throughout this correspondence. One can easily see both outer- and inner-rims exhibit zigzag pattern along the latitute direction. Under the RR, the loci of the nonhexagons, located at the corners of the rectangular patches, are shifted to the middle points of the patches. Note that while for a particular set of TCNT-defining indices, this kind of transformation may preserve the number of atoms (or triangles in the dual representation), in general this will not hold. After the transformation has been performed, the relative positions between pentagons and heptagons become staggered. And the chiral vector of that patch is changed into (1,1), which can be seen by carefully tracing the orientation of the zigzag pattern. In addition, if both IRR and ORR are enforced, they again result in eclipsed conformation, and both the inner and the outer rims of the corresponding TCNT have chiral vector (1,1), *i.e.* armchair pattern along the latitude direction.

We observe in this scheme that the issue of changing chiral vector (by ninety degrees) naturally arises as a result of IRR and ORR. To be specific, if we denote the parent TCNT by (0,0), IRR-TCNT by (1,0), ORR-TCNT by (0,1), and one with both inner and outer rims rotated by (1,1), one can easily verify that TCNT pair (0,0) and (1,1), or the other pair (0,1) and (1,0), can be viewed as belonging to the same structural type of polyhedral torus unfolded only along orthogonal orientations on graphene. This view was adopted in [1], where the authors used cSWT to represent IRR in their context. However, we believe that their terminology is incorrect, because SWT, while changes the connectivities of a trivalent carbon atom pair by rotating them for ninety degrees, does not change the number of atoms. However RR usually does change the number of atoms. In terms of the examples given in [1], this is illustrated by going from structure-type **A** ($C_{288}$) to **E** ($C_{300}$) or from **B** ($C_{300}$) to **F** ($C_{288}$), where the IRR is the operation that was performed. Actually, in certain circumstances one can apply IRR (ORR) on TCNTs belonging to $D_{nh}$ point group without changing the number of atoms, provided that their geometric indices ($n_{75}, n_{77}, n_{55}, s$) satisfy the condition $s=4k$ ($s+n_{75}=4k$). This is the reason that the numbers of atoms of **A** and **F** are the same even though they are related by ORR, since in this case this condition, $s+n_{75}=2+2=4$, holds.

**Rim rotation for $D_{nd}$-TCNTs (Stone-Wales transformation):** As for the antiprismatic case, *i.e.* TCNTs belonging to the $D_{nd}$ point group, the effects of IRR and ORR on the inner- and outer-rims are illustrated by Figure 2(a) and (b), respectively.

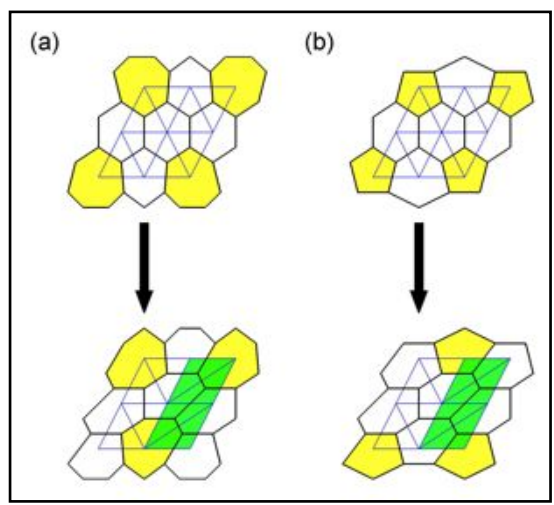

Figure 2. Illustration of rim rotation in $D_{nd}$ TCNTs. In this particular case, the effect of rim rotation is the same as that generated by the generalized Stone-Wales transformation, which conserves the number of atoms. We have shaded the transformed parts green.

Note that, in effect, only one half of the patches are transformed in this case, which is shaded in light green (Figure 2). One can find that, in analogy to the original SWT[4] which shifts the connectivities of the two selected atoms, the transformation shown in Figure 2 switches the connectivities of a whole rectangular patch of carbon atoms. Due to this fact, we have deliberately called this transformation the gSWT, g for generalized. In contrast to the RRs described above in the case of $D_{nh}$ TCNTs, gSWT does not change the number of carbon atoms. It serves as a special version of RR applied on $D_{nd}$ TCNTs for it does *rotate the rims* of the polygonal TCNT and produces TCNTs with pentagons in staggered conformation with respect to heptagons, while in a fashion different to RR defined previously[6].

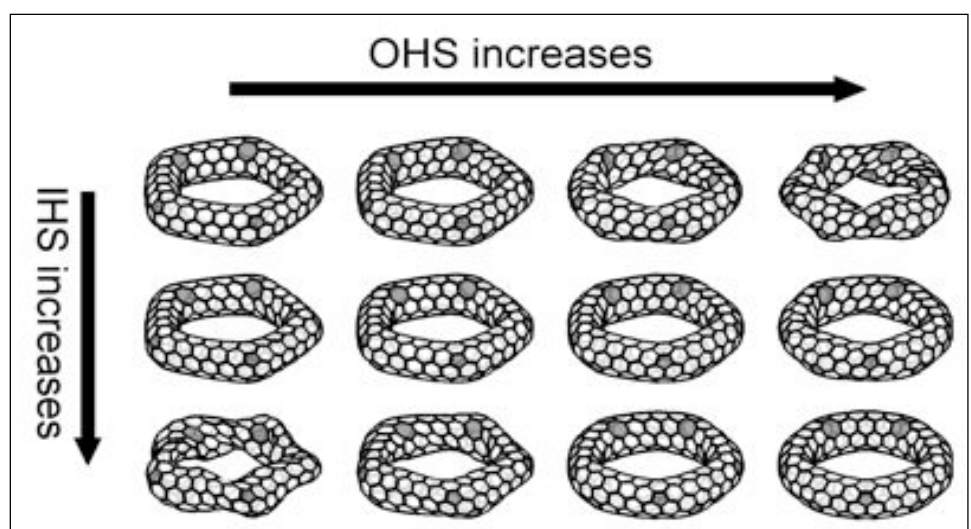

Figure 3. Horizontally shifted family of TCNTs with indices (2,1,1,4) and fivefold rotational symmetry. $D_{5h}$ and $D_{5d}$ isomers are shown at the upperleft and the lowerright corners, respectively.

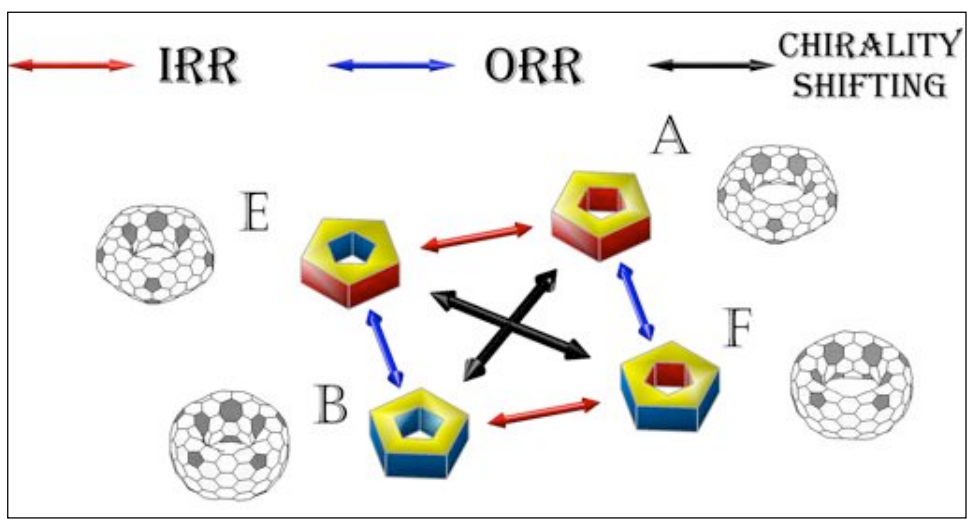

Figure 4. The relationship between four Dnh-symmetric structural types. We follow the nomenclature of [1]. The red and blue arrows refers to inner- and outer-rim rotation, respectively. The black arrows on the diagonal and antidiagonal of this square stand for the (ninety-degree) switching of chirality of the underlying graphene lattice with respect to the baseline of unfolded molecular graph.

**Horizntoal shifting:** The authors of [1] also showed that $D_{nh}$ (structure-type I and II) and $D_{nd}$ (structure-type III and IV) can be related through the appropriate HS. HS can be defined as the relative rotation of the face at the top with respect to the face at the bottom. In the process of HS, two rings of chemical bondings, selected through the intersection of TCNT and a particular horizontal plane, are broken and reformed by connecting corresponding atoms to their next-nearest neighbors. And the relative positions of the nonhexagons in the upper part of the TCNT are changed against ones in the lower part. This was first described by Berger and Avron in their seminal paper "A Classification Scheme for Toroidal Molecule"[5] on this topic. However they did not realize that by carrying out this transformation for a number of times along appropriate directions, one can eventually relate the two high-symmetric isomers ($D_{nh}$ and $D_{nd}$) in a cyclic fashion in an isomeric family of $D_n$ TCNTs. It is worth noting that this happens only if the characterizing geometric indices obey certain parity rules[2], or there will be no two high-symmetric isomers at all. Similar to the situation mentioned above, we are confused about using IRR and HS to relate the four structure-types I to IV, as revealed in Figure 3a in [1]. Since the HS is an isomerization process, structure-type **B** ($C_{300}$) and **E** ($C_{300}$) cannot possibly be transformed to **D** ($C_{288}$) and **G** ($C_{288}$), respectively, for the respective numbers of atoms are changed.

We emphasize here that HS also possesses its own subtlety. Starting from a $D_{nh}$ TCNT, where n pairs of heptagons (or pentagons) are related by having the same longitudes, we can choose either to apply HS on the inner rim or the outer rim. We denote them as inner-HS (IHS) and outer-HS (OHS), respectively[2,7]. These two choices of HS that one can perform leads to a two-dimensional series of isomeric TCNTs that can be derived from the parent $D_{nh}$ TCNT. These derived TCNTs are in general $D_n$-symmetric, and, as mentioned previously, for the set of geometric parameters of the parent TCNT obeying certain partity rules does there exist a $D_{nd}$-symmetric isomer. In fact, most of the $D_n$ TCNT isomers in this family are highly unstable except ones that are close to the two high-symmetry isomers. We shall focus on a subset of this infinite family of HS-TCNTs that can be considered relatively stable. In terms of the four geometric indices ($n_{75}, n_{77}, n_{55}, s$) characterizing the distances between nonhexagons, $D_{nh}$ TCNT and its $D_{nd}$ counterpart are related by (IHS,OHS)=($\pm s/2, \mp (n_{75}+s)/2$), where the plus and minus signs are defined by the direction of the HS chosen[2]. Thus we choose the subset of HS-TCNTs with HS parameters (IHS, OHS) lying between (0,0) and ($\pm s/2, \mp (n_{75}+s)/2$), forming an $(s/2+1)$-by-$[(n_{75}+s)/2+1]$ array of TCNTs.

Taking the $D_{5h}$-symmetric TCNT $C_{400}$, with indices (2,1,1,4), as the starting point (Figure 3). Here the parent $D_{nh}$ TCNT is shown at the upperleft corner, and its $D_{nd}$ isomer is shown at the lowerright. One notices that isomers at the upperright and the lowerleft are most strained, which correspond to (IHS,OHS)=(0,3) and (2,0) respectively. In general, isomers located at the neighborhood of the diagonal can be considered relatively stable. This implies that there might be isomers belong to $D_n$ point group, located right on the diagonal, possessing comparable stability with respect to the two highly symmetric parents. Indeed, this happens when both s and $n_{75}$ are multiples of four.

**Structural relationship:** Having these transformations defined carefully, we are now in a position to discuss the relations between the highly symmetric isomers of TCNTs in general. We follow the definition of structure-types **A-H** in [1], while taking the fivefold rotationally symmetric TCNTs as examples. Firstly, let us examine the relationships between the four $D_{5h}$-symmetric TCNTs (Figure 4).

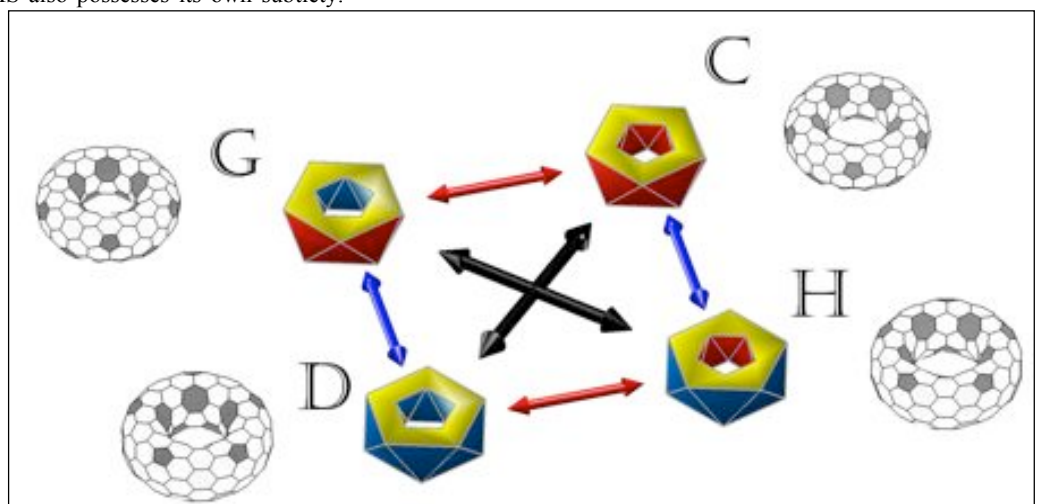

Figure 5. The relationship between the four $D_{nh}$-symmetric structure-types.

The schematic illustration of these four structure-types with $D_{nh}$ symmetry by suitable polygonal tori are shown at the central part of the figure, where their vertices are the loci of the nonhexagons. And the corresponding molecules, whose geometries were optimized by the molecular mechanics method, are shown at the two opposite sides. Starting anticlockwising from type **A**, the IRR transforms the corresponding TCNT into type **E**. Structure-type **B** is reached with ORR from type **E,** and one further IRR leads to type **F**. The cycle is completed by applying ORR again. Note that two different colors, blue and red, are used for the inner and the outer rectangular parts of the TCNT, this reflects the fact that after RR, the chirality of the "rotated" rectangular part is effectively shifted by ninety degrees, measured through the usual definition of chiral vector of straight CNTs. Thus TCNTs belonging to pairs of structure-types located at the two ends of diagonal and antidiagonal can be seen as the folding of the same kind of unfolded net, only along base lines that are orthogonal to each other. For the simplest achiral case, this corresponds to two chiral vectors (1,0) and (1,1), which are depicted in the figure as the black double-headed arrows.

One can continue this argument to the TCNTs belonging to $D_{nd}$ point group. We present the corresponding illustration in Figure 5. The four $D_{5d}$-symmetric TCNTs are related in an identical manner as in the case of $D_{5h}$ TCNTs. However, they are all isomers of graphitic toroidal $C_{240}$, in contrast to their $D_{5h}$ counterparts, which contain in general different number of carbon atoms. For the same reason, the IRR and ORR in this context are just the previously defined gSWT performed on the inner rim and the outer rim, respectively.

**Transformation cube:** Thus, we can summarize these eight structure-types **A-H** as the vertices of a cube with its edges and diagonals on the top and bottom faces representing the transformations that bring these structural types to each other as shown in Figure 6.

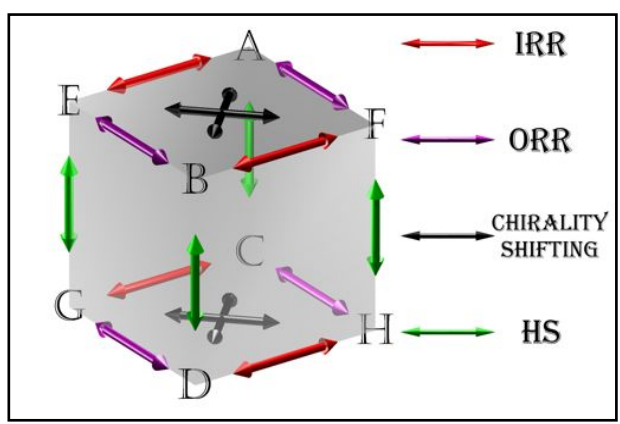

Figure 6. The cube of transformations between the eight structure-types proposed in [1]. The upper half of the cube is composed of $D_{nh}$-symmetric structure-types, and the lower half $D_{nd}$-symmetric ones.

In essence, this transformation cube is only a conceptual summary of structural types of high-symmetric TCNTs. We emphasize that for many combinations of parities of geometric indices ($n_{75}$,$n_{77}$,$n_{55}$,$s$), not all of these eight vertices of the cube exist. For example, $D_{5h}$ TCNTs **B** and **E** (which are both $C_{250}$) should be excluded from the remaining six $C_{240}$ toroidal isomers **A**,**C**, **D**, **F**, **G**, and **H**. In fact, the only possibility that all eight vertices exist is the case where both $n_{75}$ and $s$ are multiples of four, as mentioned previously. However, one can always classify prismatic/antiprismatic high-symmetric achiral TCNTs into the eight structure-types, regardless of the possible isomers located at some other corners of the cube.

**Five extra structural types:** Finally, concerning the high symmetry isomers of TCNTs, there is one particular kind of $D_{nh}$ TCNTs worth mentioning. Taking the structure-type **A** for example, n concerted gSWT can be applied to a square patch of the outer-rim, thereby switching the loci of the pentagons to the horizon of the TCNT. Similar transformation exerting on the inner rim of the elongated structure-type **F** ($C_{528}$) can lead to $D_{6h}$ TCNT isomers with six octagons lying at the horizon, formed by merging ten heptagons.

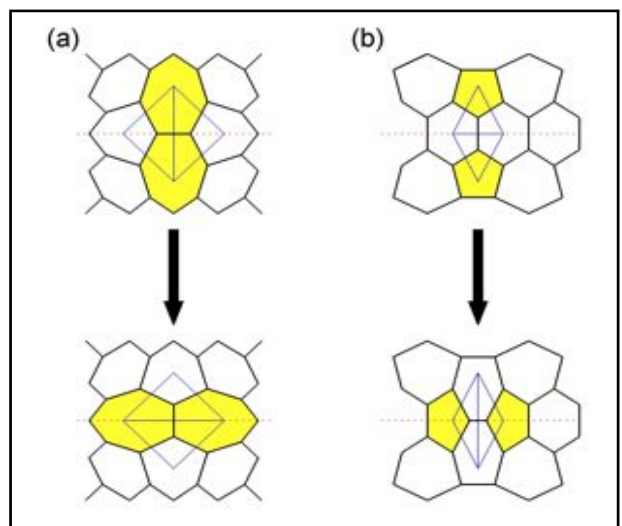

Figure 7. Illustration of GSWT applied nonhexagon pairs of $D_{nh}$-symmetric TCNTs. The dotted lines represent the inner and the outer rim equators in the respective cases.

In Figure 7(a), we focus on the inner rim of one rotational unit cell of structure-types **B** or **E**. Before the gSWT is effected, the two heptagons are related by their same longitude, where the midpoint of their respective loci is located exactly at the inner rim equator (dotted linees). After the gSWT, both of the heptagons are tranfered to the equator of the torus. The pentagons located in the outer-rim of a TCNT as shown in Figure 7(b) can be similarly transfomed. In general, a gSWT operating on the inner-rim (or outer-tim) rotates the orientation of a rhombic patch of $2n_{77}^2$ (or $2n_{55}^2$) carbon atoms. Here we have illustrated the simplest cases where $n_{77}=n_{55}=1$, with the dual of the affected atoms drawn in blue lines. Another fact worth mentioning is that although gSWT preserves the number of atoms, i.e. it is an isomerization process, it destroys the zigzag pattern along the direction of revolution, if the chiral vector of the underlying lattce is chosen to be (1,0). A clearly defined zigzag/armchair pattern throughout the torus implies that one can conceptually divide the corresponding TCNT into inter-connected trans- or cis-polyacetylene chains[8]. Among the numerous transformations mentioned above, singly RR $D_{nh}$ structure-types **E** and **F** do not possess this property as well.

We note that the gSWT described above is not applicable to all four structural types of TCNTs. In specific, gSWT of this kind can be only applied to the inner rim of structure-types **B** and **E**, and the outer rim of structure-types **A** and **E**. This leads to five more structure-types, and we summarize them in Figures 8 and 9.

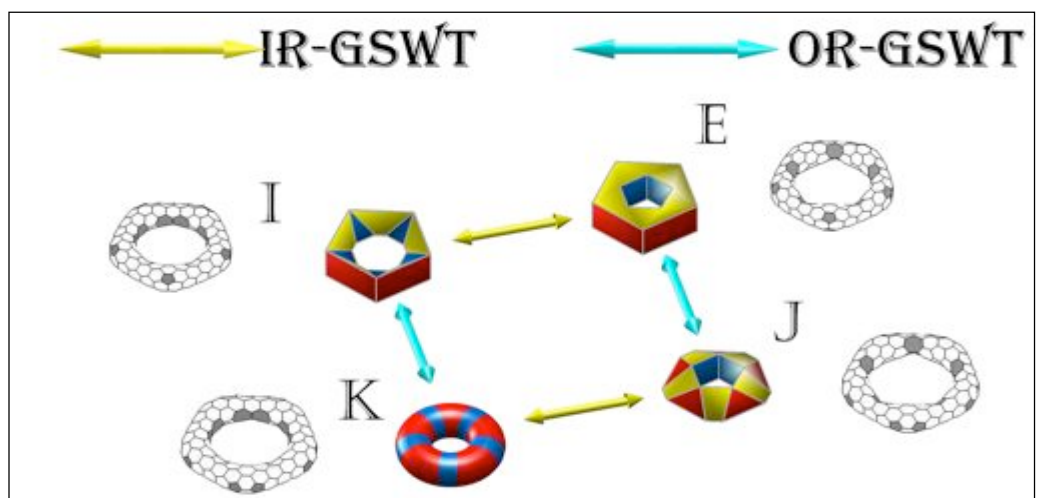

Figure 8. Three new types of $D_{nh}$-symmetric structural types derived from the GSWT operating on structure-type **E**.

Since we can apply gSWT either on the inner or the outer rim of a TCNT belonging to structure-type **E**, there are three different structure-types that can be derived from these transformations. As mentioned, gSWT, if applicable, brings the pentagons and heptagons to the respective equators of torus. For structure-types **I** and **J**, the resulting geometries of these toroidal molecules seems to be flattened along the inner and the outer equators, respectively. More intriguingly, for structure-type **K**, in which gSWT is enforced at both inner and outer rims, all of the nonhexagons are located along the equator. This leads to a family of TCNTs firstly proposed by Dunlap[9], which is formed by connecting achiral CNTs with chiral vectors (n,n) and (2n,0) alternatively for twelve times. Thus we have adopted an alternative coloring scheme to highlight this fact in Figure 8. Dunlap also generalized the idea to the connection of CNTs with arbitrary chiral vectors[10], however, they in general will be of $D_n$ symmetry, which fall out of the coverage of this correspondance.

In Figure 9, we present the rest two new structure-types **L** and **M**, corresponding to the gSWT isomers of structure-types **A** and **B**, respectively. In these two cases, further gSWT at the other rim of the torus is not allowed. Whereas for TCNTs belonging to structure-type **F**, neither their inner rims nor outer rims are applied by the gSWT. So, no further derived TCNTs can be generated.

**Summary:** We clarify the relationships between the eight structural types of TCNTs proposed in [1]. In particular, they can be identified as the eight corners of a cube of graphical transformation as shown in Figure 6. The four families with $D_{nh}$ symmetry can be related by rim rotations, and the same is true for those with $D_{nd}$ symmetries. The two sets are then connected by horizontal shiftings, thereby completing the cube. In addition, we further point out that there are five more highly symmetric $D_{nh}$ structural types that can be derived from performing the gSWT on certain TCNTs with $D_{nh}$ structural types. Particularly, one of them (structure-type **K**) is exactly the type of TCNTs first proposed by Dunlap[9,10], in which all of the nonhexagons are located at the equator of torus. This kind of TCNTs can be seen as alternatingly joining armchair and zigzag CNTs with a prescribed angle.

### Acknowledgements

We thank Prof. Y.-C. Cheng for informing us the paper by Beuerle *et al.* We also thank Prof. K.-T. Liu (NTU) and Prof. Winnie Li (NCTS, Taiwan) for useful discussions. B.Y.J. thanks Prof. R. Silbey for useful comments. We acknowledge financial support from the Center for Advanced Nano-Materials, National Taiwan University and the National Science Council, Taiwan (NSC 099-2811-M-002-166-).

**Keywords:** carbon • nanotubes • toroid topology • classification

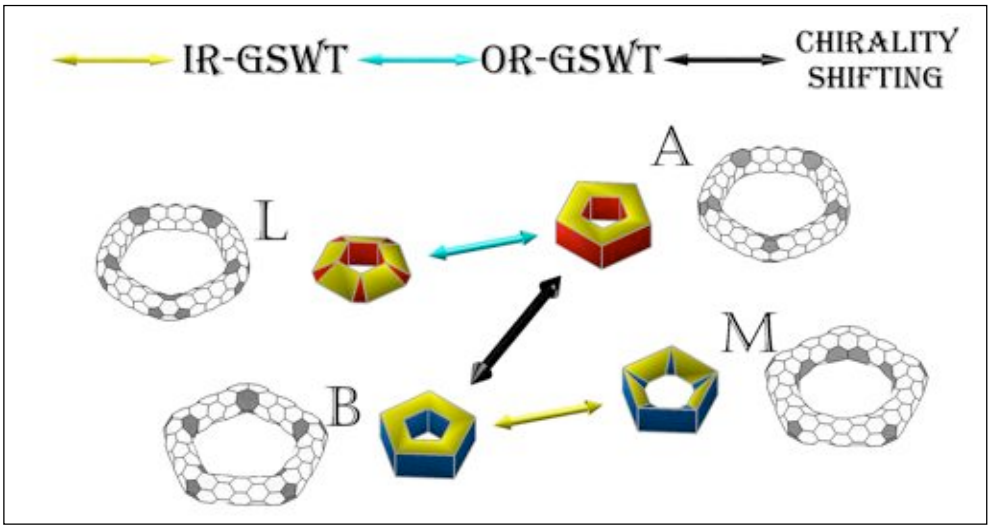

Figure 9. Two new structural types deriving from structure-types **A** and **B**.

**Eight structural types of TCNTs**
---------------

*Chern Chuang, Yuan-Chia, Fan, Bih-Yaw Jin\** ………..… Page – Page

**Comments on structural types of toroidal carbon nanotubes**

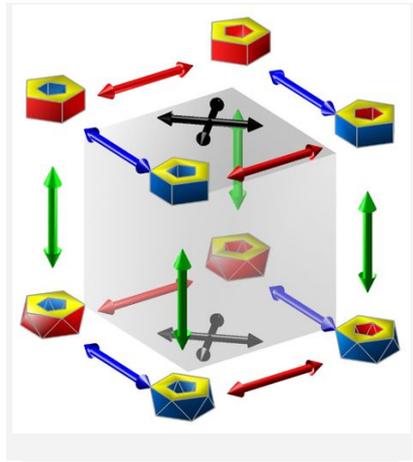

Relationships between eight structural types of toroidal carbon nanotubes can be summarized as a transformation cube with vertices denoting these distinct structures. The edges and diagonals on the top and bottom faces stand for the four different transformations that connect them.